\documentclass[11pt]{article}
\usepackage{moriond,epsfig}
\pdfoutput=1 

\usepackage{multirow,graphicx,amsmath,pst-grad,pstricks,color}

\def\Journal#1#2#3#4{{#1} {\bf #2}, #3 (#4)}

\def\PRL{\em Phys. Rev. Lett.}
\def\PL{\em Phys. Lett.}

\def\JPG{{\em J. Phys. G}}
\def\arxiv{{\em arXiv:}}

\def\be{\begin{equation}}
\def\ee{\end{equation}}
\def\bea{\begin{eqnarray}}
\def\eea{\end{eqnarray}}

\def\jpsi{J/$\psi$}
\def\psip{$\psi^{\prime}$}
\def\ups{$\Upsilon$}
\def\ee{$e^{+}e^{-}$}
\def\sqrts{$\sqrt{s_{NN}}=200$~GeV}

\begin{document}
\vspace*{4cm}
%\title{NEWS OF QUARKONIA PRODUCTION FROM THE PHENIX EXPERIMENT}
\title{RECENT QUARKONIA RESULTS FROM THE PHENIX EXPERIMENT}

\author{ZAIDA CONESA DEL VALLE FOR THE PHENIX COLLABORATION}

\address{
Laboratoire Leprince-Ringuet (\'Ecole Polytechnique, CNRS-IN2P3), Palaiseau, France
}

\maketitle\abstracts{ 
The PHENIX collaboration efforts towards constraining and understanding quarkonia production mechanisms are outlined. \jpsi~measurements and feed-down sources studies in p--p collisions at \sqrts, together with their possible implications on the hadro-production mechanisms, are discussed. 
\jpsi~photo-production in ultra-peripheral collisions at \sqrts~is also examined. 
Finally, a glimpse of the first bottomonia results in minimum bias Au--Au collisions at \sqrts~and their implications are presented. 
}

%----------------------------------------------------------------------------------------------------------------
%----------------------------------------------------------------------------------------------------------------

\section{Physics punchline}

Lattice QCD (lQCD) calculations predict that QCD matter should exist in a Quark Gluon Plasma (QGP) phase -- a state of matter where quarks and gluons are no longer confined into hadrons -- under extreme conditions of temperature and energy density ($T > 170-192$~MeV)~\cite{karsch}. 
The PHENIX experiment was designed to study QCD matter in ultra-relativistic heavy ion collisions (HIC), where such conditions are presumably achieved. 
If this matter attains a high enough temperature, lQCD predicts quarkonia bound states dissociation due to the medium color charge screening, leading to a reduction, or suppression, of quarkonia production in favor of open charm/beauty mesons~\cite{satz}.  
The experimental challenge for such studies is to separate the QGP-induced effects, observed in %the most central
 heavy-ion collisions, from the influence of the production mechanisms and usual cold-nuclear-matter effects, measured in p--p and d--Au collisions respectively. 
 
\noindent
%\jpsi~suppression has been observed both at SPS and at RHIC energies, proving the formation of a hot and dense matter~\cite{jpsi_sup_sps,jpsi_AuAu_ppg068}. 
%However, there are still some questions that remain to be answered. 
SPS and RHIC experiments have observed an anomalous \jpsi~suppression, proving the formation of a hot and dense matter in HIC~\cite{jpsi_sup_sps,jpsi_AuAu_ppg068}. However, not all aspects of their production are well understood. 
More precise data on the quarkonia states at SPS, RHIC and the LHC should %guide the theoretical interpretations. % guide the theoretical interpretations and become conclusive. 
guide the theoretical interpretations.

\noindent
Here we concentrate on the recent studies of quarkonia production mechanisms performed with the PHENIX experiment at the RHIC collider and we also comment on the latest results on bottomonia production in Au--Au collisions.

%----------------------------------------------------------------------------------------------------------------
%----------------------------------------------------------------------------------------------------------------

\section{Quarkonia production} 

There exist several theoretical attempts to describe the production mechanisms of bottomonia ($\Upsilon$, $\Upsilon^{\prime}$, $\Upsilon^{\prime\prime}$) and charmonia (\jpsi, \psip), but up to now none of them has been able to reproduce the measured differential production cross sections (rapidity and transverse momentum dependence) and polarization consistently~\cite{lansberg_review}. 
Extensive studies of quarkonia properties at different energies provide discriminating tools.

%----------------------------------------------------------------------------------------------------------------

\subsection{Charmonia constrains to the hadro-production mechanisms}
\label{sec:jpsiprod}

\begin{figure}[!htbp]
\includegraphics[width=0.44\columnwidth]{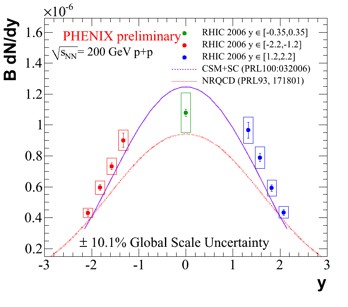}
\includegraphics[width=0.44\columnwidth]{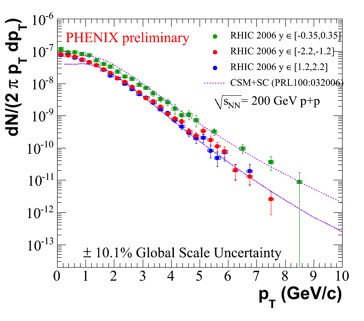}
\caption{
\jpsi~differential production cross section as a function of rapidity (left-plot) and transverse momentum (right-plot)
 in p--p collisions~\protect\cite{alex_qm} at \sqrts~compared to different theoretical calculations~\protect\cite{lansberg_jpsi_pola,nrqcd}. 
\label{fig:jpsi_xsection_y_pt}
}
\end{figure}

\begin{figure}[!htbp]
\includegraphics[width=0.4\columnwidth]{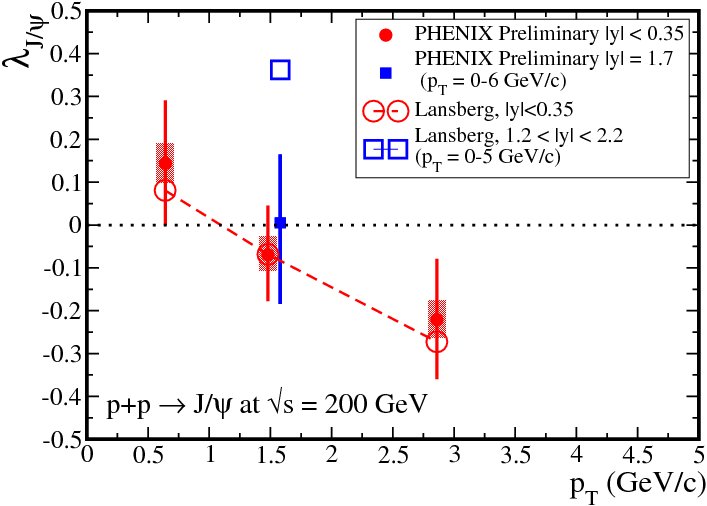}
\includegraphics[width=0.5\columnwidth]{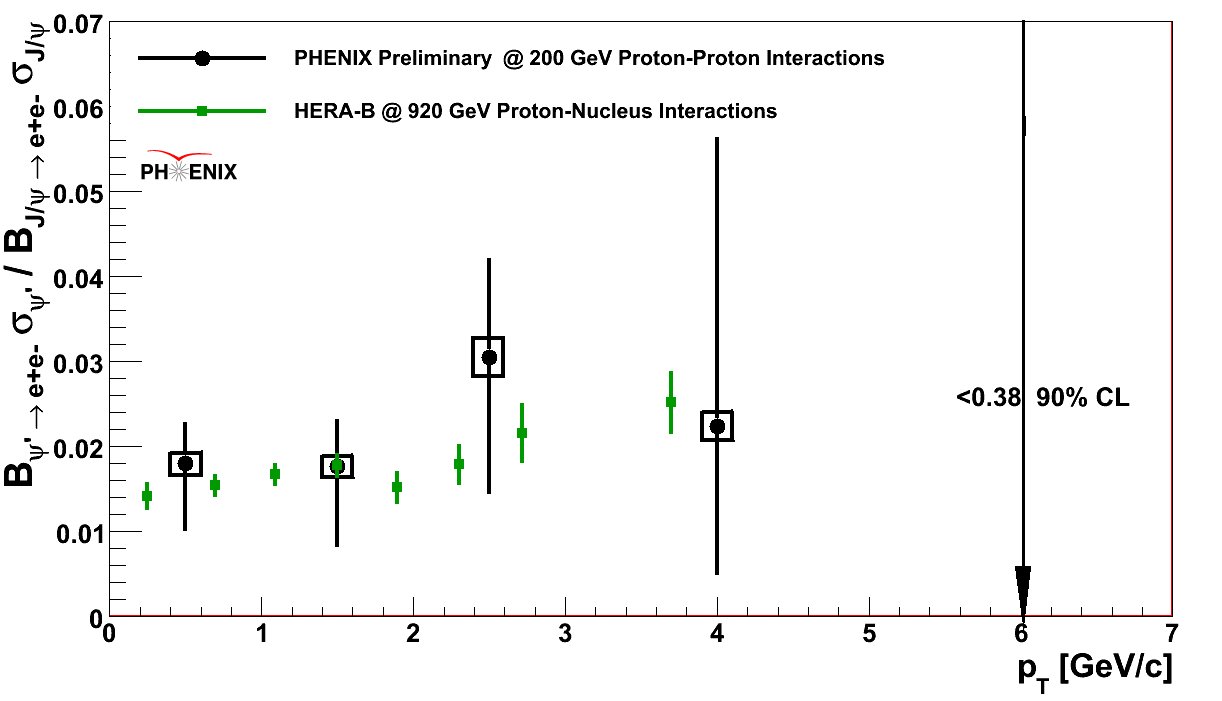}
\caption{
\jpsi~polarization (left-plot) and \psip~feed-down to \jpsi~(right-plot) in p--p collisions~\protect\cite{jpsi_psip_pol_donadelli,cesar_qm} at \sqrts~.
The polarization is drawn as a function of $y$ \& $p_t$ and compared to the theoretical calculation of~\protect\cite{lansberg_jpsi_pola}. 
\label{fig:jpsi_pol_feeddown}
}
\end{figure}

PHENIX has measured \jpsi~and \psip~production properties in p--p collisions at \sqrts. 
Fig.~\ref{fig:jpsi_xsection_y_pt} presents the \jpsi~differential production cross section~\cite{jpsi_pp_ppg069,alex_qm} rapidity ($y$) and transverse momentum ($p_t$) dependence. The absolute values for both \jpsi~and \psip~cross sections are compatible with the COM and CEM models predictions, and in agreement with the STAR collaboration results at the same center of mass energy~\cite{star_jpsi_pp}. 
The COM predictions for the \jpsi~polarization ($\lambda$) and $p_t$ dependence are challenged by the measurements~\cite{lansberg_review}, whereas the recent CSM+s-channel cut calculation~\cite{lansberg_jpsi_pola}, appropriate at intermediate $p_t$, reproduces well the $p_t$ trend and  describes satisfactorily $\lambda$ at mid-rapidity (see Fig.~\ref{fig:jpsi_pol_feeddown} left), but can not predict neither \jpsi~polarization at forward-rapidities nor the absolute value of its cross section.  
Unfortunately, the current precision on the \jpsi~rapidity dependence does not allow to rule out any hypothesis. 

\noindent
\jpsi~feed-down sources and their contributions as a function of $p_t$ also affect the theoretical calculations. 
Fig.~\ref{fig:jpsi_pol_feeddown}~right presents the PHENIX measurement of the \psip~feed-down to \jpsi, which is evaluated~\cite{jpsi_psip_pol_donadelli} to be $8.6\pm2.4\%$, approximately constant versus $p_t$, and in agreement with lower energy experiments results.  
The feed-down from $\chi_c$ decays~\cite{oda_chic} was estimated to be lower than $42\%$ with $90\%$ confidence level (C.L.), while the contribution from B decays~\cite{morino_b} is $3.6^{+2.5}_{-2.3}\%$. %, both in agreement with world wide data.

\noindent
Even though the complete interpretation of charmonia hadro-production processes is not clear yet, these measurements give additional constraints on the theoretical picture and seem to prefer the CSM+s-channel cut calculation.

%----------------------------------------------------------------------------------------------------------------
%----------------------------------------------------------------------------------------------------------------

\subsection{\jpsi~photo-production in ultra-peripheral Au--Au collisions} 
\label{sec:upc}

\begin{figure}[!htbp]
\includegraphics[width=0.5\columnwidth]{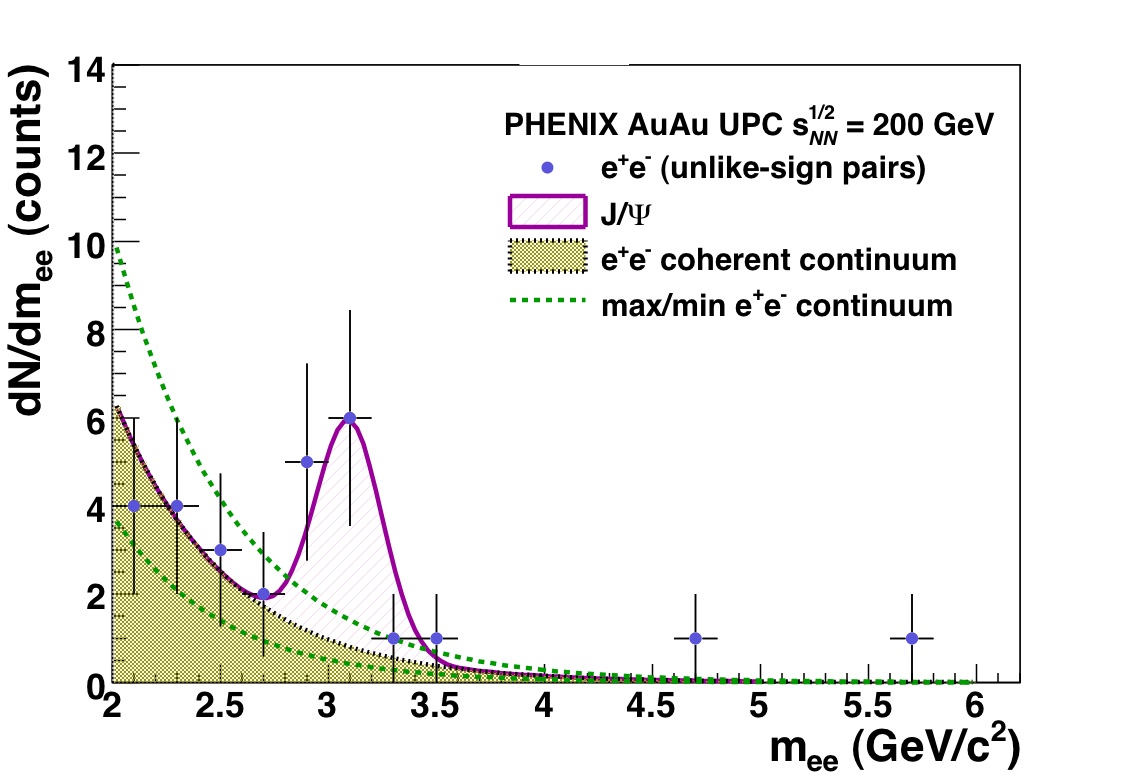}
\includegraphics[width=0.4\columnwidth, height=0.35\columnwidth]{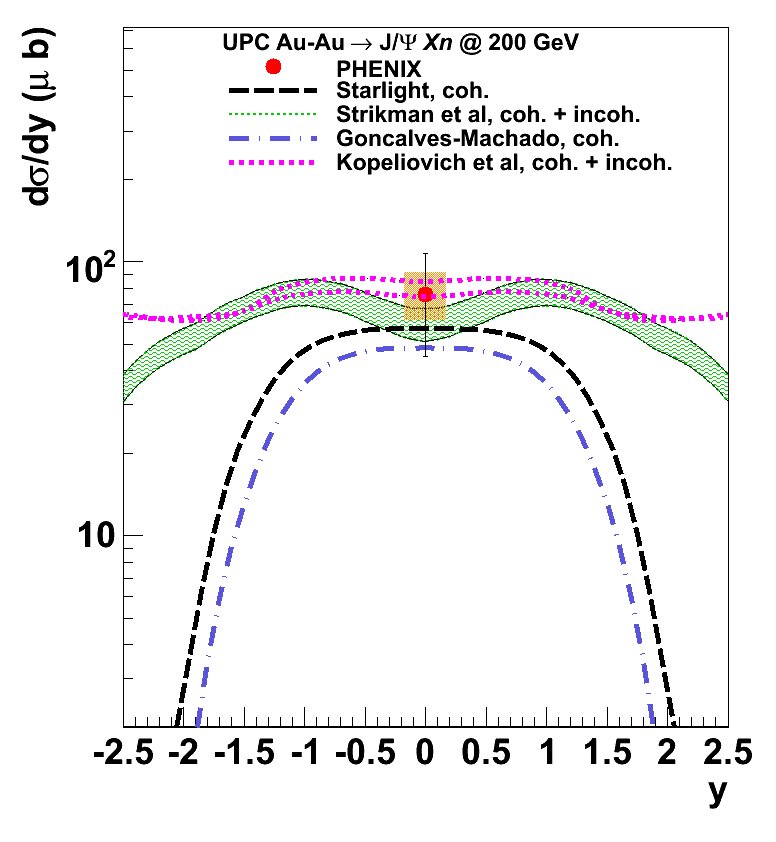}
\caption{
\jpsi~and \ee~photo-production in ultra-peripheral Au--Au collisions~\protect\cite{ppg081} at \sqrts~. 
Invariant mass spectra (left) and comparison of \jpsi~photo-production cross section to theoretical calculations (right). 
\label{fig:jpsi_upc}
}
\end{figure}

Thanks to the high photon flux generated by the ultra-relativistic moving ions, PHENIX has performed the first measurement~\cite{ppg081} ever of high-mass \ee~and \jpsi~photo-production in ultra-peripheral Au--Au collisions at \sqrts. 
%Their interest resides on testing
They allow to test the photo-production models (CSM at NLO describe well \jpsi~HERA data~\cite{lansberg_review}) and %proving the parameters involved in due to the heavy-ion origin of these interactions: the gluon-PDFs, and any PDF nuclear modifications. 
the influence of the heavy-ion origin of these interactions (the gluon PDFs and their nuclear modifications affect \jpsi~production). 
%The invariant mass spectra (Fig.~\ref{fig:jpsi_upc}~left), is constituted by 28 \ee~pairs and zero like-sign pairs with $m_{ee}> 2.0$~GeV/$c^2$. 
The invariant mass spectra is shown in Fig.~\ref{fig:jpsi_upc}~left. 28 \ee~pairs and zero like-sign pairs with $m_{ee}> 2.0$~GeV/$c^2$ are reconstructed. 
The \ee~differential production cross section at mid-rapidity, $86 \pm 23 \rm{(stat)} \pm 16 {\rm (syst)}~\mu$b/GeV/$c^2$ for $m_{ee} \in \mbox{[2.0,2.8]}$~GeV/$c^2$, is in agreement with the theoretical {\sf STARLIGHT}~\cite{starlight} LO QED calculations ($\gamma \gamma \rightarrow e^+ e^-$), suggesting the validity of this QED-interpretation on the strongly interacting regime probed by these HIC. 
The \jpsi~production cross section at mid-rapidity is $76 \pm 31 \rm{(stat)} \pm 15 {\rm (syst)}~\mu$b, and its transverse momentum dependence indicates the contribution of both coherent ($\gamma \, A \rightarrow J/\psi$) and incoherent ($\gamma \, N \rightarrow J/\psi$) production in accordance with various predictions (see refs. in \cite{ppg081}). 
Despite the large uncertainties (Fig.~\ref{fig:jpsi_upc}~right), that so far preclude any detailed conclusion on the basic ingredients of the models% (gluon-PDFs, nuclear modifications)
, this result is a proof of principle of these measurements feasibility in HIC and supports their value for such studies. % on a background-free environment. 

%----------------------------------------------------------------------------------------------------------------
%----------------------------------------------------------------------------------------------------------------

\section{Outlook, the bottomonia family} 
\label{sec:ups}

The comparison of charmonia and bottomonia production patterns is important to understand their production mechanisms and to interpret how they behave in a QGP. 
The \ups(1S) state being heavier and more tightly bound than the \jpsi, it is expected to dissociate at higher medium temperatures. Therefore, the relative patterns of the different quarkonia should provide information on the medium temperature. 

\begin{figure}[!htbp]
\includegraphics[width=0.5\columnwidth]{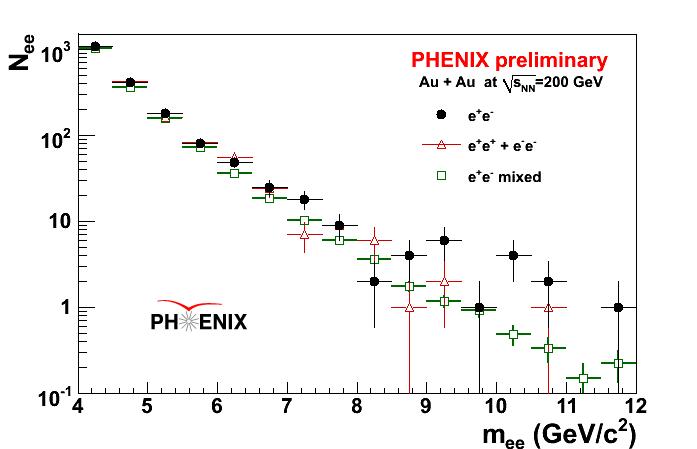}
\includegraphics[width=0.46\columnwidth]{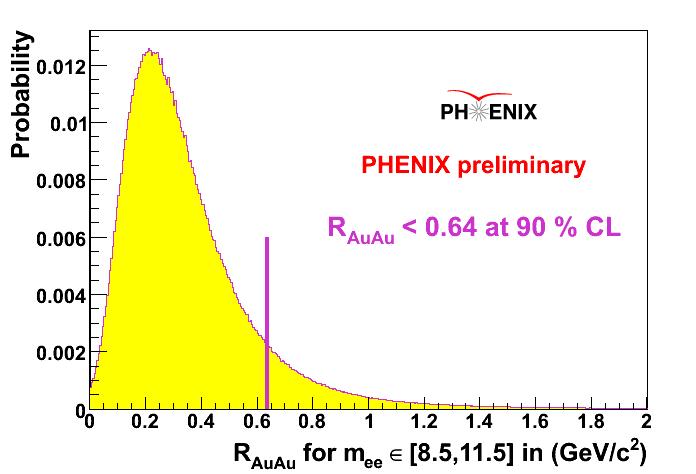}
\caption{
Left: \ee~invariant mass spectra in minimum bias Au--Au collisions at \sqrts. Right: $R_{AuAu}$, nuclear modification factor probability distribution of high mass pairs, for $m_{e^{+}e^{-}}\in [8.5,11.5]$~GeV/c$^2$.
\label{fig:ups_AuAu}
}
\end{figure}

\noindent
RHIC experiments have recently reached a high enough integrated luminosity to look at bottomonia. 
PHENIX has measured the \ups~family production cross section both at mid and forward rapidities~\cite{cesar_qm} in accordance with what was discussed in Sec.~\ref{sec:jpsiprod}. 
In addition, an excess in the \ee~invariant mass distribution, which we interpret as \ups~production (see Fig.~\ref{fig:ups_AuAu}~left), has been observed in minimum bias Au--Au collisions~\cite{ermias_qm} at \sqrts. 
Fig.~\ref{fig:ups_AuAu}~right presents the probability distribution of the high-mass nuclear modification factor $R_{AuAu}$, which corresponds to the normalized ratio of the production in Au--Au and p--p collisions. A 90\% C.L. upper-limit has been estimated to be 0.64, revealing a suppression of the high mass counts. 
Separating the continuum contribution and calculating the \ups~cross section should shed light on whether this suppression can be attributed to a suppression of the feed-down sources, of the background contamination or of the \ups~ground state.

%----------------------------------------------------------------------------------------------------------------
%----------------------------------------------------------------------------------------------------------------

\section{Highlights}

Quarkonia are hard and rare probes of the QCD matter produced in heavy-ion-collisions. 
On the one hand, their hadro-production measurements in p-p collisions seem to favor CSM calculations beyond LO, such as the CSM+s-channel cut model~\cite{lansberg_jpsi_pola}. 
While \jpsi~photo-production, observed for the first time in ultra-peripheral Au-Au collisions, is a promising tool to study its photo-production mechanisms. 
On the other hand, an excess has been observed in the \ee~invariant mass spectrum suggesting \ups~production in both p-p and minimum bias Au-Au collisions, and showing a suppression of \ee~counts within the \ups~mass region in Au-Au collisions. Further studies are needed before we can firmly conclude the implications on the \ups(1S) yields and the medium properties.

%----------------------------------------------------------------------------------------------------------------
%----------------------------------------------------------------------------------------------------------------

\end{document}